\documentclass{article}

\usepackage{arxiv}

\usepackage[utf8]{inputenc} 
\usepackage[T1]{fontenc}    
\usepackage{hyperref}       
\usepackage{url}            
\usepackage{booktabs}       
\usepackage{amsfonts}       
\usepackage{nicefrac}       
\usepackage{microtype}      
\usepackage{graphicx}
\usepackage{subfig}
\usepackage{tabularx}
\usepackage{multirow}

\graphicspath{ {./images/} }

\title{Automated Computer Program Evaluation and Projects – Our Experiences}

\author{
 Bama Srinivasan, Mala Nehru, Ranjani Parthasarathi, Saswati Mukherjee, Jeena A Thankachan \\
 Department of Information Science and Technology\\
  CEG Campus\\
  Anna University, Chennai 600 025 \\
  \texttt{Corresponding author email: bama@annauniv.edu} \\
}

\begin{document}

\maketitle

\begin{abstract}
This paper provides a few approaches to automating computer programming and project submission tasks, that we have been following for the last six years and have found to be successful. The approaches include using CodeRunner with Learning Management System (LMS) integration for programming practice and evaluation, and Git (GitHub) for project submissions and automatic code evaluation. In this paper, we describe the details of how we set up the tools and customized those for computer science courses. Based on our experiences, we also provide a few insights on using these tools for effective learning. 
\end{abstract}


\section{Introduction}
Consider a typical institution conducting a programming course (lab) for students. Here, the student is expected to have a lab notebook and prepare the assigned programming assignment each week in a well formatted structure. This structure includes an algorithm, related programming segments known as code, and the output after successful completion. Once the course instructor verifies the logic of the code, the student types it to practice programming in the lab. In the first instance, the required output may not be arrived at and the student has to patiently debug the logical and syntactical errors, which takes time. Some students readily give up and take the easiest route of writing down only the output or copying from others. Course instructors now have these tasks at hand: (i) setting up the required programming environment in labs; (ii) verifying the logic of the code that has been written by students in notebooks; (iii) helping students to debug the errors; and (iv) evaluating students’ work by checking the execution manually. Assuming a class size of 60, it is close to impossible for the course instructor to do all these tasks in the given time. 

We have addressed these shortcomings by incorporating autograding program environments, namely CodeRunner (CR) and GitHub Classroom (GHC). Both environments provide a single interface for multiple programming languages, such as C, C++, Java, and Python. Hence, a separate setup environment is not required for each of these programming labs. With CR or GHC, course instructors distribute programming assignments to students with appropriate instructions, all in virtual mode. Then students work independently based on the instructions, and the code is verified automatically. Here, there is no manual evaluation, so the course instructor does not need to go to individual students’ places and check the output of his code. The detailed work of attempts, debug methods, and marks is available to both the course instructor and the student instantly. This way, the course instructor can spend time helping students understand the programming part and debug the code if necessary. 

Students are also expected to do a team project in their course of study, where they write code in a phased manner and develop software. To encourage the collaborative environment, we use GHC for the submission of students’ code. This method provides the course instructors with information about code submissions, frequency of these submissions with changes, and the individual contribution, all in online mode. This reduces the manual time of checking the students’ progress in evaluating projects. An additional advantage is that the codes are available in a repository that is accessible to the team, even at a later point in time. 

In this paper, we provide the details of our experiences on setting up and using CR and GHC to automate the tasks of program evaluation and project submission. The paper is organized as follows. Section \ref{sec:review} discusses a few methods that are currently available for auto grading. Section \ref{sec:automate} describes the details of the installation and customization of CR. Section \ref{sec:courses} provides an outline of GHC assignment submissions and the automatic evaluation of programs. This section also includes the details of project submissions through GHC. Section \ref{sec:lesson} gives a few insights from our experiences, and Section \ref{sec:conclude} summarizes the paper. 

\section{REVIEW OF EXISTING METHODS}
\label{sec:review}
Assessment and feedback are critical aspects of the learning and teaching process \cite{benford1994ceilidh}. They not only benefit students, but also help professors identify course flaws. In programming based courses where practical skills are more important than memorization, regular practice and its evaluation are significant. Since the 1960’s, various Auto Assessment tools for automatic program evaluation in such courses have evolved to assess the code with execution (dynamic) and without execution (static) \cite{galan2019automated}. TRY is a dynamic functionality assessment tool that was implemented in 1980’s \cite{reek1989try}. Assyst, Ceilidh (also known as CourseMarker), HoGG, Online Judge, and BOSS evaluate the program’s functionality by comparing its output or checking the return values \cite{foxley1997ceilidh, foxley2001coursemaster, jackson1997grading, morris2003automatic}. WebToTeach and ELP tools evaluate single statements \cite{arnow1999webtoteach, truong2003web}. Scheme-robo and JEWL evaluate codes in Scheme and Java with graphical user interfaces, respectively \cite{english2004automated}. Automata, Dr. Scratch, EduPCR, an AA system for learning object-oriented programming, VPL, Grading Java Assignments, GradeIT, and AA for OpenGL are some of the Automatic Assessment systems that have been used in the past few years \cite{wunsche2018automatic, thiebaut2015automatic, kitaya2016online, moreno2015dr, rashid2016framework, parihar2017automatic, wang2012assessment}. 

Spanish National University of Distance Education (UNED) introduced Blocks, an open-source IDE which is limited to a few programming languages like C, C++, and Fortran \cite{codeblocks}. PASS is a web-based automated Programming Assignment assessment System, introduced and used by the Department of Computer Science, City University of Hong Kong in the year 2004; it is not open source software \cite{choy2005experiences}. Drop Project (DP) is an Auto Assessment tool used since 2018 at Lusofona University with various functionalities like standard assignment format, individual and group submissions, multiple submissions and tests, execution time and coding style \cite{cipriano2022drop}. DP is an open source tool but is limited to testing programming languages like Java, Kotlin, Clojure, and teams loginScala. CodeZinger is a proprietary tool for the educators, with costs involved for each service, and the educators need to depend on this firm for every need \cite{codezinger}. Function specific autograders like Nbgrader and jupyter autograde are restricted with jupyter notebook only, limiting them mostly to Python \cite{jupyter-autograde, nbgrader}.

CodeRunner and Virtual Programming Lab are free, open-source plugins for the Moodle learning platform \cite{coderunner}. CodeRunner, extended with OpenGL programs plugged into the Moodle learning platform, was effectively used at the University of Auckland, New Zealand, for a Computer Graphics course (200+ students). 

Git is the most prevalent version control system (VCS) nowadays \cite{git}; in the 2021 Stack Overflow survey, more than 93\% of professional developers reported using Git. As a result, there has been an increase in the use of Git in university courses. Among educators, the hosting service GitHub (GH) and its programming education tool GitHub Classroom (GHC) are particularly popular. Git workflow, autograding, offline automated processing, pull requests, admin rights to students, GitHub organization, submission dates and times, linkage with LMS like Moodle or Canvas constitute a few among many professional features \cite{githubclassroom, github}. 

From this study, it is evident that most of the tools are proprietary and therefore expensive, which does not fit the educational requirements for serving a large number of students. Some of the tools are program specific, which means they need separate resources and support for installation and customization. However, tools such as CR, GH, and the Virtual Programming Lab are free for the educational community. We have chosen CR owing to its easy customization and support. Since GH is prevalent and working with it is beneficial for students from an industry standpoint, we have incorporated GHC in our courses and programming labs.

\section{AUTOMATED CODE EVALUATION WITH CODERUNNER}
\label{sec:automate}
CodeRunner is an automated method for evaluating programming exercises. It was first developed and used by University of Canterbury as a plugin for Moodle \cite{lobb2016coderunner}. We have installed this tool in our department labs and used it to run programming courses and assessments. In this section, we provide the details of installing, customizing, and using this tool for our courses.

\subsection{Installation, Configuration and Customization}

\begin{figure} [htbp!]
    \centering
    \includegraphics{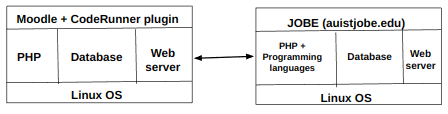}
    \caption{Setup of Moodle and CodeRunner}
    \label{fig_1}
\end{figure}

Two servers with the following configuration were assigned for the purpose of implementing Learning Management Systems: Intel Xeon E3-1200 v5, RAM of 32 GB, and Hard Disk of 16 TB. Both servers run Linux, one with CentOS and the other with Ubuntu. Moodle was installed in the former one and was used for all courses. Without disturbing this setup, we have installed the CodeRunner tool on the other server through the sandbox that supports small programming tasks, namely, Jobeserver. The design of the setup is given in Figure \ref{fig_1}. The steps of the installation of Moodle initially were carried out with Apache, MySQL, and PHP \cite{moodle}. For the CR, the Jobeserver was installed first, along with Apache and the related programming packages such as Python 3, C, C++, Java, and Octave \cite{jobe}. Since it is a private server, the API authentication mechanism was not installed. Once installed, this server was connected to the Moodle server through its IP address or domain address. For our setup, we have configured the domain address as "auistjobe.edu."

By default, the address of the plugin is Canterbury web server, namely ‘jobe2.cosc.Canterbury.ac.nz’. This however is not a good option because: (i) the number of students accessing CR from our department is more than 100 and (ii) the server has to be connected to the internet if default settings are used. These two limitations are addressed by having our own sandbox installed in a separate server as mentioned earlier. 

\begin{figure}[hhtbp!]
    \centering
    \includegraphics[width=0.8\linewidth]{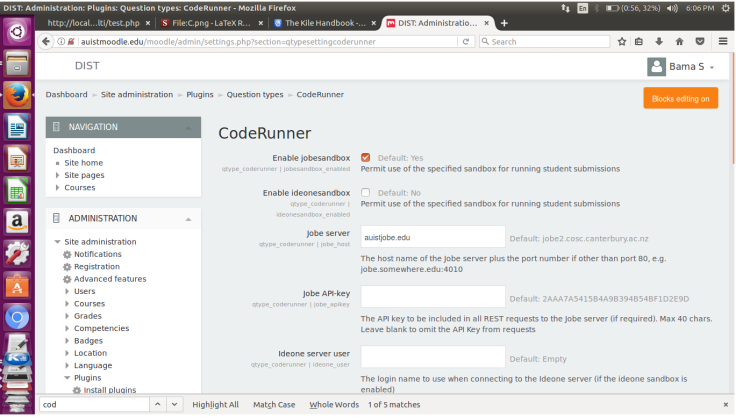} 
    \caption{Customization through domain address}
    \label{fig_2}
\end{figure}

From Moodle administration, first the plugin of CR was installed. Then in the configuration option, the domain address of the machine with Jobeserver was given, which is shown in Figure \ref{fig_2}. 

We have further customized the whole setup by having two domain addresses - one which can be operated only within the labs without internet connection and the other, which can be accessed anywhere in the department/ university with internet connectivity. This customization helps the teachers to load questions in Moodle from their place and students can answer those questions within the lab premises without internet connectivity as a closed book assessment.

\subsection{Code Runner Usage}

CR supports multiple languages such as C, C++, Java, Python, PHP and Octave. Once the plugin of CR is installed, the question bank options from Moodle has CR, through which programming questions can be added as an adaptive mode.

\begin{figure}[htbp!]
    \centering
    \includegraphics [width=0.6\linewidth]{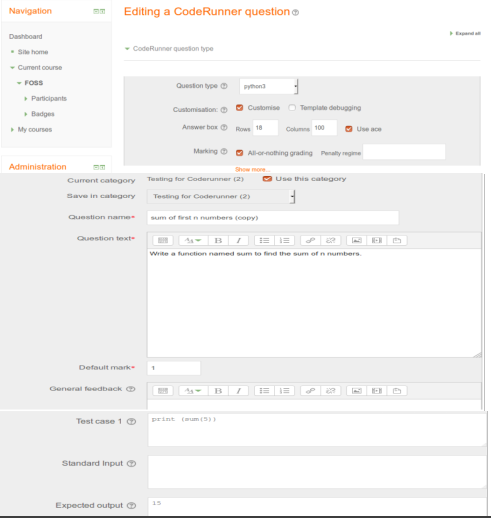}
    \caption{CodeRunner - Adding Question }
    \label{fig_3}
\end{figure}

The teacher has the flexibility of providing the customized or debugging options with the required test cases. Figure \ref{fig_3} shows these options for a sample question of finding the sum of first \textit{n} natural numbers for the programming language Python with a test case. Here, any number of test cases can be added with this setup.

\begin{figure}[h]
 \centering
 \subfloat[Correct Answer] {\includegraphics [width=0.5\linewidth]{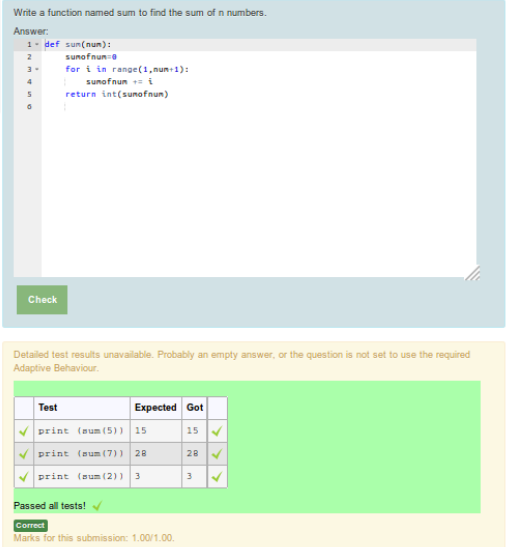}\label{fig_4}}
 \subfloat[Wrong Answer] {\includegraphics [width=0.5\linewidth]{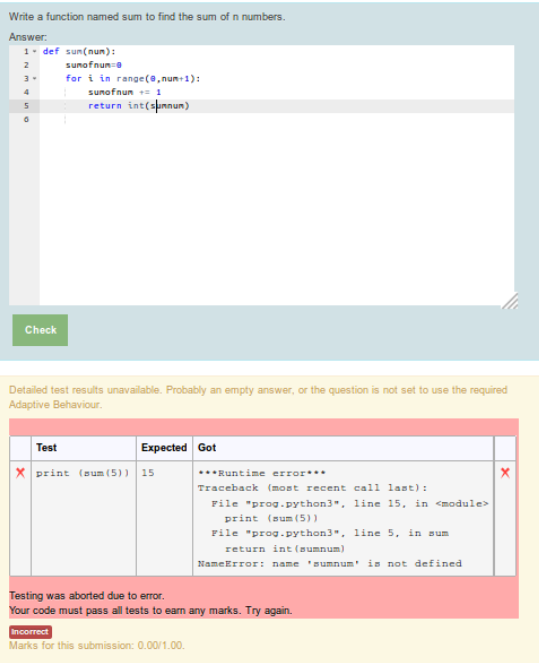}\label{fig_5}}
\caption{Evaluation with CR}
\label{fig:eval}
\end{figure}

With these questions in the question bank, the teacher can provide a quiz from the course activity option in Moodle as an adaptive mode. Students can view the question and code within the window and debug with the check option. Figure \ref{fig:eval}  indicates the attempt of students for correct and wrong answers.

We have conducted programming assessments for multiple languages to more than 450 students simultaneously using this setup. Students can debug the code and work until they get the correct answer according to the test cases. Teachers also get the immediate feedback and marks of each student. 

While CR is used within our University settings, we also explored GHC for online programming tasks. This is described in the next section.

\section{USE OF GITHUB FOR PROGRAMMING COURSES AND PROJECTS}
\label{sec:courses}
Git is a free and open source distributed version control system used by software developers for development of software \cite{git}. GitHub (GH) is an internet hosting service for software development based on Git \cite{github}. It provides extensive support with free resources for teachers and students through GitHub Classroom (GHC) \cite{githubclassroom}. In the next section, we describe the details of setting up classrooms for our courses. 

\subsection{Creation of Classrooms through GHC}

\begin{figure}[h]
    \centering
    \includegraphics [width=0.8\linewidth]{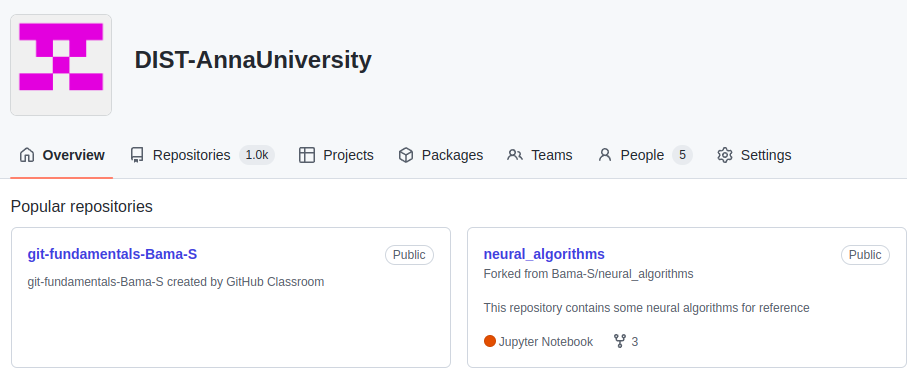} 
    \caption{Creation of Organization in GH }
    \label{fig_6}
\end{figure}

To set up a classroom, first we created an organization namely ‘DIST-AnnaUniversity’ where the repositories of students and teachers could be accommodated, as shown in Figure \ref{fig_6}. 

\begin{figure}[h]
    \centering
    \includegraphics [width=0.8\linewidth]{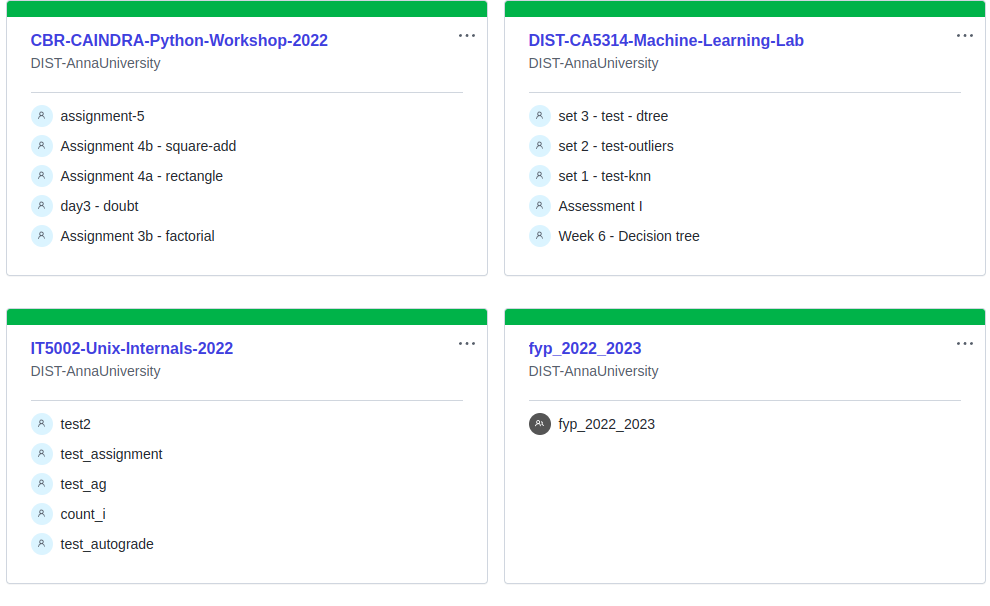} 
    \caption{Classrooms created through GHC }
    \label{fig_7}
\end{figure}

An academic teacher enrolled through GH can create multiple organizations after due verifications from GH. Then we added multiple classrooms, each for a different course linked with the organization that was created earlier. A snapshot of sample classrooms – ‘Python Workshop, Machine Learning Lab, Unix Internals and fyp 2022 2023’ from our department is shown in Figure \ref{fig_7}. Next, we enrolled students in the specific classroom. GHC provides two methods for enrolment (i) through LMS such as Moodle (ii) through a manual procedure. Since we had Moodle set up for the course, we connected it with GHC. More than one teacher could be added for a single classroom, which can be done using the “TA and admins” menu. 

With this setup in place assignments can be added. We have experimented with two different types of assignment- (i) Programming Courses and laboratories (ii) Project Submissions. These are described in the next two sections. 

\subsection{GHC for Programming Courses and Laboratories}

The assignment section of GHC is customized for the programming courses in terms of creation of template repository, assigning each student an individual repository and adding relevant modules and test cases for autograding. These steps are explained below. 

\subsubsection{Creation of Template with starter code}

As an initial step, a repository has to be created in an organization as a template. As a running example, here we provide the description of how we created a Python programming exercise with the autograding method for Machine Learning Course. The exercise is to fit a straight line using Linear Regression between the two attributes of the dataset of iris \cite{iris_dataset}. The actual question that we used is given below. 

\begin{itemize}
    \item[a)] Find the average petal width in the program \texttt{average.py}. (5 points)
    \item[b)] Find the range of petal width in the program \texttt{range.py}. (7 points)  
    \item[c)] In the program \texttt{regression.py}, fit a straight line between petal length and petal width. Suppose the equation is $y = ax + b$, find the variables $a$ and $b$ first. Then for a new petal width value of 5.3, find the predicted value of $y$. (13 points)
\end{itemize}

For this specific question, we first created three empty program files namely, ‘average.py’, ‘range.py’ and ‘regression.py’ with the required Python modules. In each of the programs, hints in the form of comments were given, which help the student to add the lines of code at the specific places. For example, the question of regression.py appears as follows. Here, each student can use a different logic to find the values of ‘a’ and ‘b’, using which the new predicted value with the variable ‘new pl’ can be determined. 

\paragraph{Question:}
In this program, you will fit a straight line between petal length and petal width. Suppose the equation is y = ax+b, find the variables ‘a’ and ‘b’ first. Then for a new petal width value of 5.1, find the predicted value of y.

\begin{verbatim}
import pandas as pd 
import numpy as np 
from sklearn.model_selection import train_test_split 
from sklearn.linear_model import LinearRegression 
from sklearn import metrics 

# read csv file 
# Take X as petal length 
# Take y as petal width 
# Split train and test with 30% and set random state as 0 
# use regression from scikit learn 
# find b and round off to 2 decimal places 
# find a and round off to 2 decimal places 

new_pw = 5.3  # new x value 

# calculate y
new_pl = print(round(new_pl, 2))  # round off to 2 decimal places 
\end{verbatim}

Similar comments were given for program files ‘average.py’ and ‘range.py’. 

Thus, the question set for this exercise consists of (i) the dataset ‘iris.csv’ (ii) three python programming files with comments and (iii) readme.md consisting of questions. 

\begin{figure}[h]
    \centering
    \includegraphics [width=0.6\linewidth]{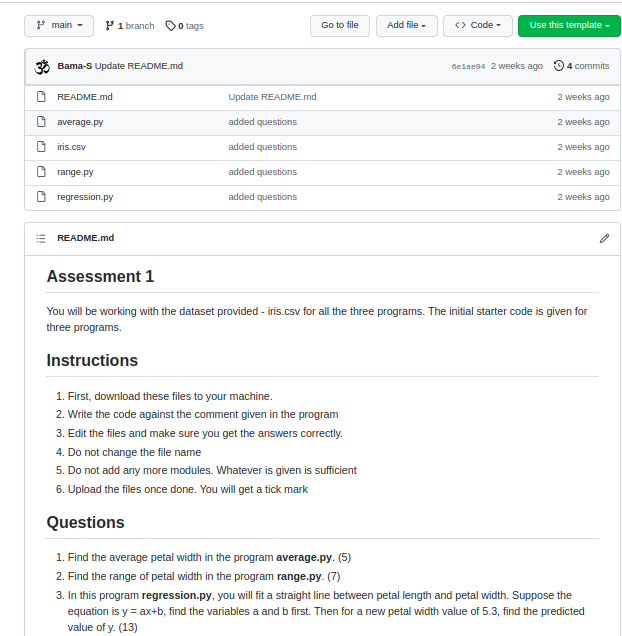}
    \caption{Creation of a template repository }
    \label{fig_8}
\end{figure}

In the GHC organization ‘DIST-AnnaUniversity’ (Figure \ref{fig_6}), a repository namely, ‘ml-assessment1-2022’ for this exercise was created first. The above question set was then pushed/ uploaded and the repository was set as template, as shown in Figure \ref{fig_8}. This template was then linked to the assignment section of GHC, which is described next. 

\subsubsection{Adding Assignment in GHC}

Within each classroom, a new assignment can be created with the options of name, deadline, individual /group assignment, repository visibility with options of private/ public, admin access methods, support of template and editors, addition of test cases and pull request. For the above programming exercise, we have included the following details. 

\begin{itemize}
\item Name as `Assessment I'
\item Deadline of 90 minutes, which can be altered based on the needs
\item Individual assignment, since each student has to work independently
\item Private repository
\item Provide admin access, so that students have full ownership
 \item Template link of `ml-assessment1-2022', which was created as described in Section IV-B1.
 \item GHC provides codespaces for education purposes, which help students to code within a browser. Hence, we enabled the editor of Codespaces which supports vscode [37].
\item Addition of test cases, which is a mandatory step for autograding programming exercises. This is detailed in Section IV-B3
\item Pull request, so that teachers can give immediate feedback
\end{itemize}

\begin{table}[h]
\caption{Test Case Settings for the three programs}
\begin{tabular}{|p{2.4 cm}|p{4.5 cm}|p{4.5 cm}|p{5 cm}|}
    \hline
    \textbf{Test Name} & \textbf{Average} & \textbf{Range} & \textbf{Regression} \\
    \hline
   Setup Command & \texttt{sudo -H pip3 install pandas;} & \texttt{sudo -H pip3 install pandas;}& \texttt{sudo -H pip3 install pandas;}  \\
   $~$ &$~$ & $~$ & \texttt{sudo -H pip3 install scikit-learn;} \\
   $~$ & $~$ & $~$ & \texttt{sudo -H pip3 install numpy;} \\
    \hline
    Run command & \texttt{python3 average.py} & \texttt{python3 range.py} & \texttt{python3 regression.py} \\
    \hline
    Expected Output & 1.2 & 2.4 & 1.85 \\
    \hline
     Comparison & included & included & included \\
     \hline
     Timeout & 10 & 10 & 10 \\
     \hline
     Points & 5 & 7 & 13 \\
     \hline
\end{tabular}
\end{table}

\subsubsection{Addition of test cases}

\begin{figure}[htbp]
    \centering
    \includegraphics [width=0.5\linewidth]{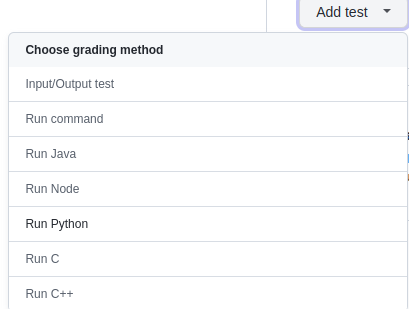}
    \caption{Adding tests for autograding}
    \label{fig_9}
\end{figure}

GHC provides autograding facility through the addition of GH actions. In the assignment options, we can add test case as ‘input/output’, ‘run command’ or add specific unit tests for languages python, Java, Node, C and C++, as shown in Figure \ref{fig_9}. 

For the above example, we used ‘input/output’ option with the settings shown in Table I. The different settings are described below. 

\begin{itemize}
\item {The name of the test, which can be identified distinctly for the three different programs.}
\item {Setup commands are essential, where Python libraries are to be included for the execution of programs. The program 'regression.py' (listed above) specifies the libraries of pandas, sklearn, and numpy, which need to be installed before execution of the program. Hence, the setup command should include all the necessary packages required for the program execution.}
\item {Run command is for execution of Python programs. Since all the three Python programs are of version 3, 'python3 <filename>' is used.}
\item {Expected output of each program is to be given. In the running example, the output values are 1.2, 2.4, and 1.85 for average.py, range.py, and regression.py, respectively.}
\item {The comparison has three options: (i) included, (ii) exact, and (iii) regex. Here, we have given 'included' because this particular test case will check for the value which is included in the print statement.
\item Default timeout of 10 minutes is given.}
\item {Marks (points) for the specific question as 5, 7, and 13 are given for 'average.py', 'range.py', and 'regression.py', respectively.}
\end{itemize}

Alternatively, ‘Run Python’ option (Figure \ref{fig_9}) can also be used, through which ‘pytest’ is incorporated by GHC. In such cases, the settings and the question type varies. We have also experimented with this type and have found it be useful.  

\begin{figure}[htbp]
    \centering
    \includegraphics [width=0.8\linewidth]{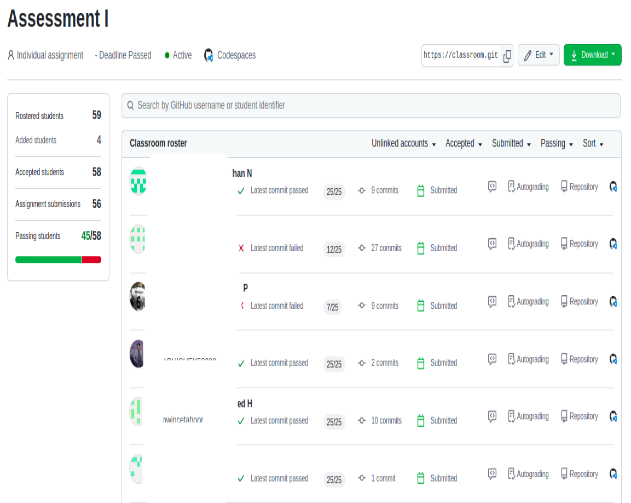} 
    \caption{Consolidated student view of assignments }
    \label{fig_10}
\end{figure}

With these settings, the assignment can be created and the link can be distributed to students. Once the student takes up the assignment, the instructor can immediately view the status of students in terms of the number of students who have accepted, submitted and passed with their scores and a link to their repository, as shown in Figure \ref{fig_10}. It can be seen that one gets points, only if the test case succeeds in each of the questions. The instructor can view each student’s repository to verify the programming structure for the correctness. If the feedback is to be given, it can be done through pull request. 

\begin{figure}[htbp]
    \centering
    \includegraphics [width=0.8\linewidth]{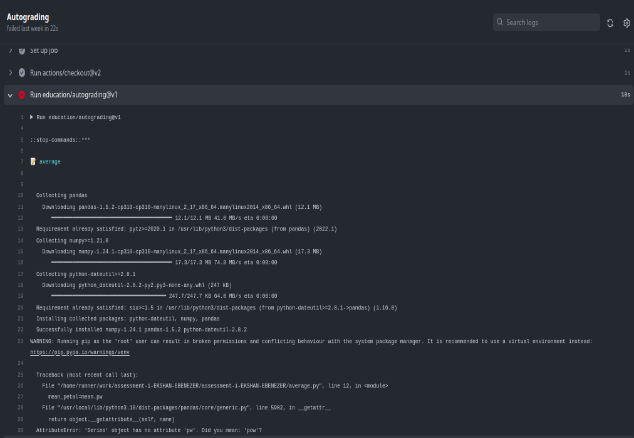} 
    \caption{Code execution at the background }
    \label{fig_11}
\end{figure}

\begin{figure}[htbp]
    \centering
    \includegraphics [width=0.8\linewidth]{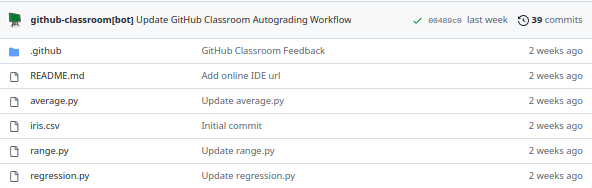} 
    \caption{Successful completion }
    \label{fig_12}
\end{figure}

All the above details are described only from the teacher’s viewpoint. From the students’ perspective, once he/ she accepts the assignment, a unique repository is created by GHC in the same organization with the template where questions are loaded. The student can then download those to their machine, work on it and after ensuring the correctness can upload/ push to their repository. GHC automatically executes the code in the background and checks with the available test cases. This is visible to the student and can be used to debug the code if it fails, as shown in Figure \ref{fig_11}. If all the test cases pass, the repository has a green tick mark indicating that the student has successfully completed the exercise, as shown in Figure \ref{fig_12}. 

\begin{figure}[htbp]
    \centering
    \includegraphics [width=0.8\linewidth]{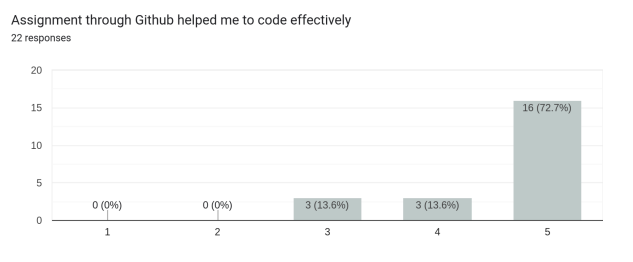} 
    \caption{Response of participants from Python Workshop}
    \label{fig_13}
\end{figure}

Apart from programming assessments, we also conduct online and hybrid programming workshops through GHC. Recently, we conducted a Python workshop for beginners spanning six continuous days with two hours’ online sessions each day for five days and a single day in offline mode. At the beginning of the workshop, participants were neither aware of coding nor comfortable with GH practices, but were able to code and submit programming exercise in GHC within 3 days. Through this method, not only participants get benefited with the coding part, the instructor also can clear the doubts by going through their work and providing the feedback instantly and correcting the codes. The response from participants for this course is given in Figure \ref{fig_13}, where 1 means least-effective and 5 means most-effective. 

\subsubsection{GHC for Project submission}

\begin{figure}[htbp]
    \centering
    \includegraphics [width=0.8\linewidth]{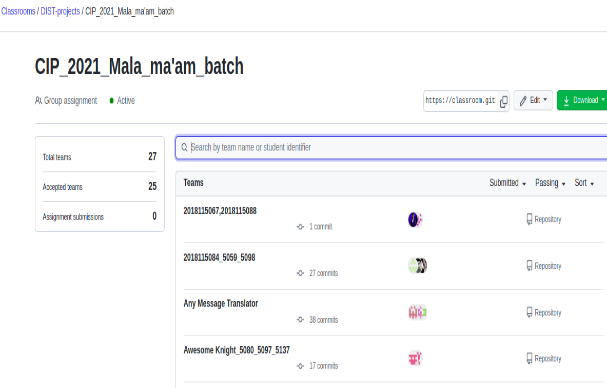} 
    \caption{Submission of team projects }
    \label{fig_14}
\end{figure}

\begin{figure}[htbp]
    \centering
    \includegraphics [width=0.8\linewidth]{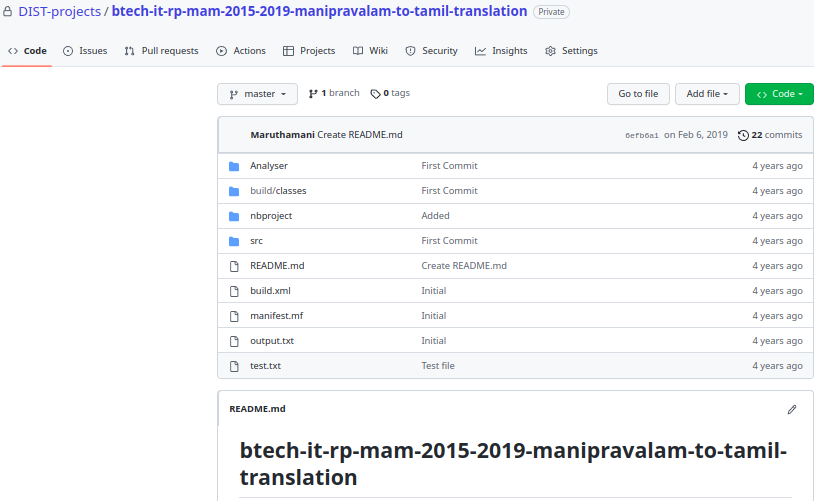} 
    \caption{Sample FYP team repository in GHC }
    \label{fig_15}
\end{figure}

Projects that involve code submissions are encouraged to be submitted through GHC. We have used GHC for project courses such as Final Year Project (FYP), Creative Innovative Project, Socially Relevant Project and Programming with Open Source. For these courses, we use a similar set up of GHC and assignment that was described above, but in a collaborative mode as a group assignment listing the number of members for each team. A sample course with 27 teams is shown in Figure \ref{fig_14}. Students, project guide and staff in-charge of project/ course are added as collaborators. While students can work on the code and push these to this repository in a collaborative environment, the staff can verify and provide feedback instantaneously. The project staff in charge can use the number of commits, contributors and frequency of code submission information to evaluate the project. A sample repository of a FYP is shown in Figure \ref{fig_15}, indicating the number of commits and the information of codes.

Thus with GHC, we have automated programming exercises and project submission tasks for the last six years. In the next section, we provide a few insights based on our experiences. 

\section{LESSONS LEARNT}
\label{sec:lesson}

Although there are challenges in adapting to the virtual environment, it turns out to be beneficial in the long run. Here, we project some pointers which may help other educational institutions to move towards autograding mechanisms. 

\begin{itemize}
\item CR is free and open source with zero pricing, making it ideal for educational settings.
\item GHC, presently maintained by Microsoft, provides limitless private repositories along with other resources for the academic community. This benefits students by providing an industry-ready environment for practicing computer programming and contributing to open-source projects.
\item Initially, students may have a learning curve to get used to CR and GHC, but after two to three attempts, students find it easier and more useful.
\item For closed-book assessments, CR in an intranet setup without an internet connection fares better.
\item For regular programming exercises and open-book assessments, GHC works well due to the method of question-wise test cases, automated grading, and instantaneous feedback from course instructors to students.
\item Both CR and GHC support multiple programming languages and can be used to conduct programming assessments and labs in parallel.
\item GHC is suited for both in-person labs and online sessions.
\item Students find this method of programming useful as it enables them to participate in programming contests and industry-oriented internships and jobs.
\item Training of faculty is essential for both CR and GHC usage in programming courses.
\item With GHC, there is always the possibility of copying among students. To eliminate this practice, free plagiarism detection tools such as MOSS can be used. Another approach adopted was having random questions distributed to students through Moodle quiz, significantly reducing the chances of copying.
\end{itemize}

\section{CONCLUSION}
\label{sec:conclude}

In this paper, we have described a few automation mechanisms for our programming labs and project submissions. The mechanisms include CR and GHC. We have customized CR based on our course requirements and the major advantage is that a closed book programming assessment could be automated with this setup. We have used GHC for both learning as well as assessments in programming courses through online mode.  This automation has multiple advantages from the perspective of students and course instructors. From the student point of view, they can understand the programming constructs better, practice at their own pace and get a real time experience of contribution to the open source community, especially when they use GHC. Instructors save time by eliminating manual evaluation and at the same time help students in learning.

\bibliographystyle{unsrt}  
\bibliography{references}  

\begin{thebibliography}{10}

\bibitem{benford1994ceilidh}
SD~Benford, EK~Burke, E~Foxley, and CA~Higgins.
\newblock Ceilidh: A courseware system for the assessment and administration of
  computer programming courses in higher education.
\newblock In {\em Proceedings of the Interdisciplinary Workshop on Complex
  Learning in Computer Environments}, 1994.

\bibitem{galan2019automated}
Daniel Galan, Ruben Heradio, Hector Vargas, Ismael Abad, and Jose~A Cerrada.
\newblock Automated assessment of computer programming practices: The 8-years
  uned experience.
\newblock {\em IEEE Access}, 7:130113--130119, 2019.

\bibitem{reek1989try}
Kenneth~A Reek.
\newblock The try system-or-how to avoid testing student programs.
\newblock In {\em Proceedings of the twentieth SIGCSE technical symposium on
  Computer science education}, pages 112--116, 1989.

\bibitem{foxley1997ceilidh}
Eric Foxley, Colin Higgins, Edmund Burke, Cleveland Gibbon, and Abdullah~Mohd
  Zin.
\newblock The ceilidh system an overview and some experiences of use.
\newblock In {\em Asian Technology Conference in Mathematics, ATCM97, Penang,
  Malaysia}, pages 430--441, 1997.

\bibitem{foxley2001coursemaster}
Eric Foxley, Colin Higgins, Pavlos Symeonidis, and Athanasios Tsintsifas.
\newblock The coursemaster automated assessment system--a next generation
  ceilidh.
\newblock In {\em Computer assisted assessment workshop}, 2001.

\bibitem{jackson1997grading}
David Jackson and Michelle Usher.
\newblock Grading student programs using assyst.
\newblock In {\em Proceedings of the twenty-eighth SIGCSE technical symposium
  on Computer science education}, pages 335--339, 1997.

\bibitem{morris2003automatic}
Derek~S Morris.
\newblock Automatic grading of student's programming assignments: an
  interactive process and suite of programs.
\newblock In {\em 33rd Annual Frontiers in Education, 2003. FIE 2003.},
  volume~3, pages S3F--1. IEEE, 2003.

\bibitem{arnow1999webtoteach}
David Arnow and Oleg Barshay.
\newblock Webtoteach: an interactive focused programming exercise system.
\newblock In {\em FIE'99 Frontiers in Education. 29th Annual Frontiers in
  Education Conference. Designing the Future of Science and Engineering
  Education. Conference Proceedings (IEEE Cat. No. 99CH37011}, volume~1, pages
  12A9--39. IEEE, 1999.

\bibitem{truong2003web}
Nghi Truong, Peter Bancroft, and Paul Roe.
\newblock A web based environment for learning to program.
\newblock In {\em Proceedings of the 26th Australasian computer science
  conference-Volume 16}, pages 255--264. Citeseer, 2003.

\bibitem{english2004automated}
John English.
\newblock Automated assessment of gui programs using jewl.
\newblock {\em ACM SIGCSE Bulletin}, 36(3):137--141, 2004.

\bibitem{wunsche2018automatic}
Burkhard~C W{\"u}nsche, Zhen Chen, Lindsay Shaw, Thomas Suselo, Kai-Cheung
  Leung, Davis Dimalen, Wannes van~der Mark, Andrew Luxton-Reilly, and Richard
  Lobb.
\newblock Automatic assessment of opengl computer graphics assignments.
\newblock In {\em Proceedings of the 23rd annual ACM conference on innovation
  and technology in computer science education}, pages 81--86, 2018.

\bibitem{thiebaut2015automatic}
Dominique Thi{\'e}baut.
\newblock Automatic evaluation of computer programs using moodle's virtual
  programming lab (vpl) plug-in.
\newblock {\em Journal of Computing Sciences in Colleges}, 30(6):145--151,
  2015.

\bibitem{kitaya2016online}
Hiroki Kitaya and Ushio Inoue.
\newblock An online automated scoring system for java programming assignments.
\newblock {\em International journal of information and education technology},
  6(4):275, 2016.

\bibitem{moreno2015dr}
Jes{\'u}s Moreno-Le{\'o}n, Gregorio Robles, and Marcos Rom{\'a}n-Gonz{\'a}lez.
\newblock Dr. scratch: Automatic analysis of scratch projects to assess and
  foster computational thinking.
\newblock {\em RED. Revista de Educaci{\'o}n a Distancia}, (46):1--23, 2015.

\bibitem{rashid2016framework}
Nuraini~Abdul Rashid, Lau~Wei Lim, Ooi~Sin Eng, Tan~Huck Ping, Zurinahni
  Zainol, and Omar Majid.
\newblock A framework of an automatic assessment system for learning
  programming.
\newblock In {\em Advanced Computer and Communication Engineering Technology:
  Proceedings of ICOCOE 2015}, pages 967--977. Springer, 2016.

\bibitem{parihar2017automatic}
Sagar Parihar, Ziyaan Dadachanji, Praveen~Kumar Singh, Rajdeep Das, Amey
  Karkare, and Arnab Bhattacharya.
\newblock Automatic grading and feedback using program repair for introductory
  programming courses.
\newblock In {\em Proceedings of the 2017 ACM conference on innovation and
  technology in computer science education}, pages 92--97, 2017.

\bibitem{wang2012assessment}
Yanqing Wang, Hang Li, Yuqiang Feng, Yu~Jiang, and Ying Liu.
\newblock Assessment of programming language learning based on peer code review
  model: Implementation and experience report.
\newblock {\em Computers \& Education}, 59(2):412--422, 2012.

\bibitem{codeblocks}
{Code Blocks}: Open source, cross platform, free c, c++ and fortran ide.
\newblock \url{http://www.codeblocks.org/}.
\newblock Accessed: Jul. 25, 2019.

\bibitem{choy2005experiences}
Marian Choy, U~Nazir, Chung~Keung Poon, and Yuen-Tak Yu.
\newblock Experiences in using an automated system for improving students’
  learning of computer programming.
\newblock In {\em Advances in Web-Based Learning--ICWL 2005: 4th International
  Conference, Hong Kong, China, July 31-August 3, 2005. Proceedings 4}, pages
  267--272. Springer, 2005.

\bibitem{cipriano2022drop}
Bruno~Pereira Cipriano, Nuno Fachada, and Pedro Alves.
\newblock Drop project: An automatic assessment tool for programming
  assignments.
\newblock {\em SoftwareX}, 18:101079, 2022.

\bibitem{codezinger}
{Code Zinger}.
\newblock \url{https://codezinger.com/pricing/}.

\bibitem{jupyter-autograde}
{Jupyter autograde}.
\newblock \url{https://pypi.org/project/jupyter-autograde/}.

\bibitem{nbgrader}
{NB grader}.
\newblock \url{https://nbgrader.readthedocs.io/en/stable/}.

\bibitem{coderunner}
{Coderunner}.
\newblock \url{https://moodle.org/plugins/qtype_coderunner}.

\bibitem{git}
{Git}.
\newblock \url{https://git-scm.com/}.
\newblock Accessed: 01-04-2024.

\bibitem{githubclassroom}
{GitHub Classroom}.
\newblock \url{https://classroom.github.com/}.

\bibitem{github}
{GitHub}.
\newblock \url{https://github.com/}.

\bibitem{lobb2016coderunner}
Richard Lobb and Jenny Harlow.
\newblock Coderunner: A tool for assessing computer programming skills.
\newblock {\em ACM Inroads}, 7(1):47--51, 2016.

\bibitem{moodle}
{Moodle}.
\newblock \url{https://moodle.org/}.
\newblock 12-12-2023.

\bibitem{jobe}
{Jobe}: A cross-platform job control manager and automation engine.
\newblock \url{https://github.com/trampgeek/jobe}.

\bibitem{iris_dataset}
Iris dataset.
\newblock \url{https://archive.ics.uci.edu/ml/datasets/iris}.
\newblock Accessed: 15-12-2023.

\end{thebibliography}



\end{document}